\documentclass[reprent,aps,prb,twocolumn,
superscriptaddress,nofootinbib]{revtex4-2}
\usepackage{amsmath,amssymb}
\usepackage{graphicx}
\usepackage{bm}
\usepackage{multirow}
\usepackage{hyperref}
\usepackage{braket}

\usepackage{color}

\allowdisplaybreaks[2]

\begin{document}

\title{
Hybrid skyrmion and anti-skyrmion phases in polar $C_{\rm 4v}$ systems
}
\author{Satoru Hayami}
\email{hayami@phys.sci.hokudai.ac.jp}
\affiliation{Graduate School of Science, Hokkaido University, Sapporo 060-0810, Japan}

\begin{abstract}
We investigate the stability of the skyrmion crystal phase in a tetragonal polar system with the Dzyaloshinskii-Moriya interaction by focusing on the symmetry of ordering wave vectors forming the skyrmion crystal. 
Our analysis is based on numerical simulations for an effective spin model, which is derived from the weak-coupling regime in the Kondo lattice model on a polar square lattice.
We show that a hybrid square skyrmion crystal consisting of Bloch and N\'eel spin textures emerges even under polar $C_{\rm 4v}$ symmetries when the ordering wave vectors correspond to low-symmetric wave vectors in momentum space, which is in contrast to the expectation from the Lifshitz invariants. 
We also show the instability toward the anti-skyrmion crystal and rhombic skyrmion crystal depending on the direction of the Dzyaloshinskii-Moriya vector in momentum space. 
Furthermore, we show that the regions of the skyrmion crystal phases are affected by taking into account the symmetric anisotropic exchange interaction. 
Our results open the potential direction of engineering the hybrid skyrmion crystal and anti-skyrmion crystal phases in polar magnets. 
\end{abstract}

\maketitle

\section{Introduction}

Spatial inversion symmetry is one of the important factors in determining physical properties in solids. 
When the spatial inversion symmetry is broken, the system acquires various properties like chirality and polarity, which become the origin of parity-breaking phenomena, such as the Edelstein effect~\cite{edelstein1990spin, Yip_PhysRevB.65.144508, Fujimoto_PhysRevB.72.024515,yoda2018orbital, Massarelli_PhysRevB.100.075136}, nonlinear Hall effect~\cite{Sodemann_PhysRevLett.115.216806, tokura2018nonreciprocal, Nandy_PhysRevB.100.195117, Xiao_PhysRevB.102.024109}, and piezoelectric effect~\cite{curie1894symetrie}. 
Such breaking of the spatial inversion symmetry also leads to exotic states of matter, such as odd-parity multipole orderings~\cite{Spaldin_0953-8984-20-43-434203, Yanase_JPSJ.83.014703, Hayami_PhysRevB.90.081115, hitomi2014electric, Hayami_doi:10.7566/JPSJ.84.064717, Fu_PhysRevLett.115.026401, hayami2016emergent, Hayami_PhysRevLett.122.147602} and unconventional superconductors~\cite{Micklitz_PhysRevB.80.100506, Sato_PhysRevB.81.220504, Wei_PhysRevB.89.014506, Kozii_PhysRevLett.115.207002, Venderbos_PhysRevB.94.180504, Nakamura_PhysRevB.96.054501, Ruhman_PhysRevLett.118.227001, Ishizuka_PhysRevB.98.224510, Yatsushiro_PhysRevB.105.155157}. 
In this way, noncentrosymmetric systems provide a fertile platform to explore attracting quantum states and their related physical properties in condensed matter physics. 

The lack of spatial inversion symmetry often affects the stability of magnetic phases in magnetic materials. 
The most familiar example is the Dzyaloshinskii-Moriya (DM) interaction that originates from the relativistic spin--orbit coupling~\cite{dzyaloshinsky1958thermodynamic,moriya1960anisotropic}. 
The DM interaction tends to favor the single-$Q$ spiral spin configuration by combining the ferromagnetic exchange interaction. 
It also becomes the origin of multiple-$Q$ spin configurations, which are expressed as a superposition of multiple spiral waves. 
Especially, a skyrmion crystal (SkX), which is characterized by a multiple-$Q$ state, emerges by further considering the effect of an external magnetic field~\cite{Bogdanov89, Bogdanov94,rossler2006spontaneous}. 
The SkX has been extensively studied in both theory and experiments~\cite{Muhlbauer_2009skyrmion,yu2010real, yu2011near, seki2012observation, Adams2012, nagaosa2013topological}, since it exhibits not only parity-breaking physical phenomena but also topological ones, such as the topological Hall effect~\cite{Lee_PhysRevLett.102.186601, Neubauer_PhysRevLett.102.186602, Hamamoto_PhysRevB.92.115417}. 

A variety of SkXs have been so far found in noncentrosymmetric magnets, which are classified into Bloch SkXs, N\'eel SkXs, and anti-SkXs depending on the sign of the topological charge and helicity of skyrmion~\cite{Tokura_doi:10.1021/acs.chemrev.0c00297}. 
From the energetic viewpoint, their emergence is expected from the Lifshitz invariants that correspond to the energy contribution by the DM interaction~\cite{dzyaloshinskii1964theory, kataoka1981helical, Bogdanov89, Bogdanov94}. 
Since the form of the Lifshitz invariants is determined by the crystallographic point-group symmetry, one can find what types of SkXs are realized once the symmetry of the materials is identified.  
For example, the Bloch SkXs appear in the chiral point groups~\cite{Muhlbauer_2009skyrmion,yu2010real, yu2011near, seki2012observation, Adams2012, fujishiro2019topological, hayami2021field}, the N\'eel SkXs appear in the polar point group~\cite{kezsmarki_neel-type_2015, bordacs2017equilibrium, Fujima_PhysRevB.95.180410, Kurumaji_PhysRevLett.119.237201}, and anti-SkXs appear in the point groups $D_{\rm 2d}$ and $S_4$~\cite{nayak2017discovery, Huang_PhysRevB.96.144412, peng2020controlled, karube2021room}. 
Furthermore, the hybrid SkX, which is characterized by a superposition of Bloch- and N\'eel-type windings, has been identified in synthetic multilayer magnets~\cite{legrand2018hybrid, li2019anatomy, Liyanage_PhysRevB.107.184412}. 

In the present study, we investigate the possibility of the emergent hybrid SkX and anti-SkX under polar symmetry, which are not expected from the Lifshitz invariants. 
By focusing on the symmetry of ordering wave vectors constituting the SkX, we find that the instability toward such SkXs is brought about by the DM vector lying on the low-symmetric wave vectors, which has been recently observed in EuNiGe$_3$~\cite{singh2023transition, matsumura2023distorted}.
We demonstrate that such a situation naturally happens in the Kondo lattice model with the antisymmetric spin--orbit coupling (ASOC) on a polar square lattice, where long-range Ruderman-Kittel-Kasuya-Yosida (RKKY) interaction~\cite{Ruderman,Kasuya,Yosida1957} plays an important role. 
Then, we construct the magnetic phase diagrams in a wide range of model parameters by performing the simulated annealing for an effective spin model with momentum-resolved DM interaction. 
We show that three types of SkXs are realized in an external magnetic field depending on the direction of the DM vector: square SkX (S-SkX), rhombic SkX (R-SkX), and anti-SkX. 
In the S-SkX, the constituent ordering wave vectors are orthogonal to each other, while they are not in the R-SkX and anti-SkX. 
Moreover, we find that the induced SkXs are characterized as the hybrid SkXs to have both Bloch and N\'eel spin textures. 
We also discuss the effect of symmetric anisotropic exchange interaction on the SkX phases. 
The present results provide another possibility of material design in terms of the SkXs by taking into account the symmetry of the ordering wave vectors. 

The rest of this paper is organized as follows. 
In Sec.~\ref{sec: Effective spin interactions in itinerant electron systems}, we introduce the Kondo lattice model in a tetragonal polar system and derive the RKKY interaction. 
We show that there is a directional degree of freedom in terms of the DM vector at low-symmetric wave vectors. 
In Sec.~\ref{sec: Effective spin model and method}, we construct an effective spin model and outline numerical simulated annealing used to investigate the ground-state phase diagram. 
Then, we show the instability toward three types of SkXs in Sec.~\ref{sec: Skyrmion crystal phases}. 
We examine the effect of additional magnetic anisotropy on the stability of the SkX in Sec.~\ref{sec: Effect of symmetric anisotropic exchange interaction}. 
We summarize the results of this paper in Sec.~\ref{sec: Summary}.

\section{Effective spin interactions in itinerant electron systems}
\label{sec: Effective spin interactions in itinerant electron systems}

Let us start with the Kondo lattice model on a two-dimensional square lattice under the $C_{4v}$ point group, which consists of the itinerant electrons and classical localized spins~\cite{Meza_PhysRevB.90.085107, Hayami_PhysRevLett.121.137202}. 
The Hamiltonian is given by 
\begin{eqnarray}
\label{eq: Ham_krep}
\mathcal{H}=& &\sum_{\bm{k} \sigma} (\varepsilon_{\bm{k}}-\mu) c^{\dagger}_{\bm{k}\sigma}c_{\bm{k}\sigma} 
+ J_{\rm K} \sum_{\bm{k}\bm{q}\sigma\sigma'}
c^{\dagger}_{\bm{k}\sigma}\bm{\sigma}_{\sigma \sigma'}c_{\bm{k}+\bm{q}\sigma'} \cdot \bm{S}_{\bm{q}} \nonumber \\
&+&  \sum_{\bm{k}} \bm{g}_{\bm{k}} \cdot c^{\dagger}_{\bm{k}\sigma}\bm{\sigma}_{\sigma \sigma'}c_{\bm{k}\sigma'},  
\end{eqnarray}
where $c^{\dagger}_{\bm{k}\sigma}$ and $c_{\bm{k}\sigma}$ are the creation and annihilation operators of an itinerant electron at wave vector $\bm{k}$ and spin $\sigma$, respectively. 
$\bm{S}_{\bm{q}}$ represents the Fourier transform of a localized spin $\bm{S}_i$ at site $i$ with the fixed length $|\bm{S}_i|=1$.
The first term represents the hopping term of itinerant electrons, where $\varepsilon_{\bm{k}}$ is the energy dispersion and $\mu$ is the chemical potential. 
We take $\varepsilon_{\bm{k}}=-2t_1 (\cos k_x + \cos k_y) - 4t_2 \cos k_x \cos k_y$ with the nearest-neighbor hopping $t_1$ and next-nearest-neighbor hopping $t_2$; we set the lattice constant of the square lattice as unity and choose $t_1=1$ and $t_2=-0.8$, although the choice of the hopping parameters does not affect the following results at the qualitative level.  
The second term stands for the Kondo coupling between itinerant electron spins and localized spins, where $J_{\rm K}$ is the exchange coupling constant and $\bm{\sigma}=(\sigma^x,\sigma^y,\sigma^z)$ is the vector of Pauli matrices. 
The third term stands for the Rashba ASOC that originates from the spin--orbit coupling under polar symmetry; $\bm{g}_{\bm{k}} = \alpha (\sin k_y,-\sin k_x)=-\bm{g}_{-\bm{k}}$; $\alpha$ is the amplitude of the ASOC. 

By supposing the situation where $J_{\rm K}$ is small enough compared to the bandwidth of itinerant electrons, we derive the effective spin Hamiltonian in the weak-coupling region. 
Within the second-order perturbation in terms of $J_{\rm K}$, the spin Hamiltonian is given by~\cite{Hayami_PhysRevLett.121.137202} 
\begin{align}
\label{eq:RKKYHam}
\mathcal{H}^{\rm RKKY} =-J_{\rm K}^2  \sum_{\bm{q},\nu,\nu'}
\chi^{\nu\nu'}(\bm{q})
S^\nu_{\bm{q}} S^{\nu'}_{-\bm{q}},  
\end{align}
where $\nu, \nu'=x,y,z$. 
$\chi^{\nu\nu'}(\bm{q})$ with $\bm{q}=(q_x, q_y)$ represents the spin-dependent magnetic susceptibility of itinerant electrons, which depends on the hopping parameters, ASOC, and the chemical potential. 
Under the $C_{\rm 4v}$ symmetry, nonzero components in $\chi^{\nu\nu'}(\bm{q})$ are generally given by~\cite{Yambe_PhysRevB.106.174437} 
\begin{align}
\label{eq: suscep}
\chi(\bm{q})= \left(
\begin{array}{ccc}
{\rm Re}[\chi^{xx}(\bm{q})] & {\rm Re}[\chi^{xy}(\bm{q})] & -i{\rm Im}[\chi^{zx}(\bm{q})]\\
{\rm Re}[\chi^{xy}(\bm{q})] & {\rm Re}[\chi^{yy}(\bm{q})] & i{\rm Im}[\chi^{yz}(\bm{q})] \\
i {\rm Im}[\chi^{zx}(\bm{q})] & -i{\rm Im}[\chi^{yz}(\bm{q})] & {\rm Re}[\chi^{zz}(\bm{q})]
\end{array}
\right),
\end{align} 
where $\chi^{\nu \nu'}(\bm{q})={\rm Re}[\chi^{\nu \nu'}(\bm{q})]+{\rm Im}[\chi^{\nu \nu'}(\bm{q})]$. 
The effective interaction $J_{\rm K}^2 \chi^{\nu\nu'}(\bm{q})$ corresponds to the $\bm{q}$ component of the generalized RKKY interaction~\cite{shibuya2016magnetic, Hayami_PhysRevLett.121.137202}; the antisymmetric imaginary components in $\chi^{\nu \nu'}(\bm{q})$ correspond to the DM interaction, while the symmetric real components correspond to the isotropic and anisotropic exchange interactions. 
The DM vector at $\bm{q}$ is given by $\bm{D}_{\bm{q}}=J^2_{\rm K}({\rm Im}[\chi^{yz}(\bm{q})], {\rm Im}[\chi^{zx}(\bm{q})])$. 

\begin{figure}[tb!]
\begin{center}
\includegraphics[width=1.0\hsize]{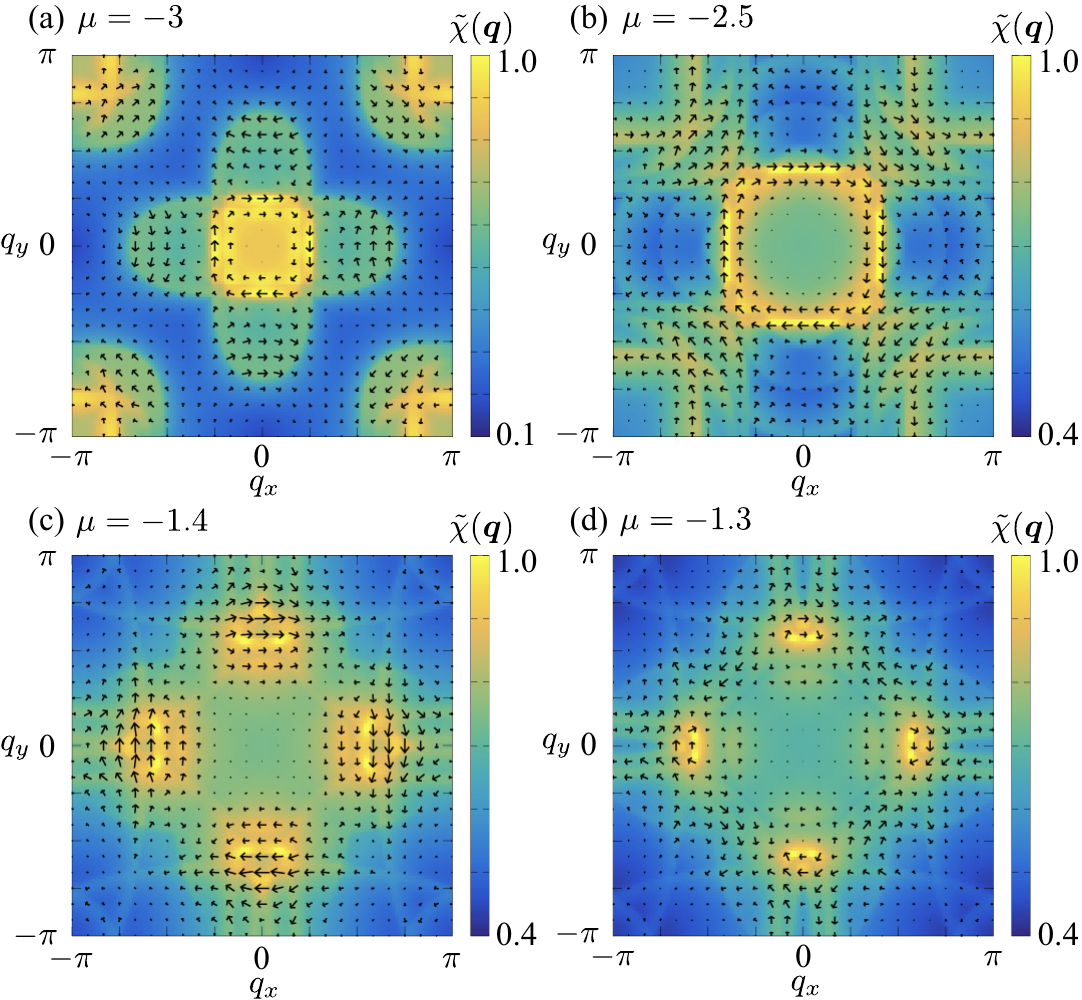} 
\caption{
\label{fig: chi} 
Contour plots of the normalized bare susceptibility $\tilde{\chi}(\bm{q})=\lambda(\bm{q})/\lambda^{\rm max}$ derived from the Kondo lattice model in Eq.~(\ref{eq: Ham_krep}) with $t_1=1$, $t_2=-0.8$, and $\alpha=0.5$ at (a) $\mu=-3$, (b) $\mu=-2.5$, (c) $\mu=-1.4$, and (d) $\mu=-1.3$. 
The wave vectors that give $\lambda^{\rm max}$ are (a) $\bm{q}=(0, \pi/4)$, (b) $\bm{q}=(2\pi/5, 3\pi/20)$, (c) $\bm{q}=(\pi/12,11\pi/20)$, and (d) $\bm{q}=(\pi/20, 17\pi/30)$ and their symmetry-related wave vectors. 
The arrows represent the direction of the DM vector at each wave vector, whose lengths stand for the magnitude of the DM interaction. 
}
\end{center}
\end{figure}

The magnetic instability of the spin model in Eq.~(\ref{eq:RKKYHam}) occurs at the wave vector that gives the maximum eigenvalue of $\chi(\bm{q})$ in Eq.~(\ref{eq: suscep}). 
We show the contour plot of the largest eigenvalues for magnetic susceptibility in each $\bm{q}$, $\lambda(\bm{q})$, at $\alpha=0.5$ for several $\mu$ in Fig.~\ref{fig: chi}; we set $\mu=-3$ in Fig.~\ref{fig: chi}(a), $\mu=-2.5$ in Fig.~\ref{fig: chi}(b), $\mu=-1.4$ in Fig.~\ref{fig: chi}(c), and $\mu=-1.3$ in Fig.~\ref{fig: chi}(d). 
We take the grids of $\bm{k}$ and $\bm{q}$ are $2400^2$ and $120^2$, respectively.  
We normalize the magnetic susceptibility as $\tilde{\chi}(\bm{q})=\lambda(\bm{q})/\lambda^{\rm max}$, where $\lambda^{\rm max}$ represents the largest eigenvalues for all $\bm{q}$. 
$\tilde{\chi}(\bm{q})$ exhibits the maximum value at $\bm{q}=(0, \pi/4)$ in Fig.~\ref{fig: chi}(a), $\bm{q}=(2\pi/5, 3\pi/20)$ in Fig.~\ref{fig: chi}(b), $\bm{q}=(\pi/12,11\pi/20)$ in Fig.~\ref{fig: chi}(c), and $\bm{q}=(\pi/20, 17\pi/30)$ in Fig.~\ref{fig: chi}(d). 
It is noted that $\tilde{\chi}(\bm{q})$ becomes maximum at the other wave vectors that are connected to the above wave vectors by the rotational and/or mirror symmetries under the point group $C_{\rm 4v}$. 
For example, $\tilde{\chi}(\bm{q})$ also becomes maximum at $\bm{q}=(\pi/4,0)$ in the case of Fig.~\ref{fig: chi}(a), while $\tilde{\chi}(\bm{q})$ becomes maximum at $\bm{q}=(-3\pi/20,2\pi/5)$ and $\bm{q}=(3\pi/20,2\pi/5)$ in the case of Fig.~\ref{fig: chi}(b). 
The spiral state with these ordering wave vectors is chosen as the ground state. 

In the following, we focus on the behavior of the imaginary part of the magnetic susceptibility, which corresponds to the DM interaction. 
We show the momentum-resolved DM vectors $\bm{D}_{\bm{q}}/J^2_{\rm K}=({\rm Im}[\chi^{yz}(\bm{q})], {\rm Im}[\chi^{zx}(\bm{q})])$ as the arrows in Fig.~\ref{fig: chi}, where the length and direction of the arrows express the magnitude and direction of the DM vectors, respectively. 
One finds that the direction of the DM vectors in each wave vector is fixed to the direction perpendicular to $\hat{\bm{q}}_z\times \bm{q}$ ($\hat{\bm{q}}_z$ represents the unit vector along the $q_z$ direction) when $\bm{q}$ lies on the high-symmetric $\langle 100 \rangle$ and $\langle 110 \rangle$ lines, while it is arbitrary for the other $\bm{q}$. 
This is attributed to the presence of the mirror plane on the high-symmetric $\langle 100 \rangle$ and $\langle 110 \rangle$ lines, which imposes the constraint on the direction of the DM vector. 

The above result indicates that the spiral plane realized in the ground state depends on the position of ordering wave vectors that give the maximum magnetic susceptibility. 
When the ordering wave vectors lie on the high-symmetric $\langle 100 \rangle$ and $\langle 110 \rangle$ lines as found in Fig.~\ref{fig: chi}(a), the spiral plane is parallel to $\bm{q}$; the cycloidal spiral state becomes the ground state, which is expected from the Lifshitz invariants under the $C_{\rm 4v}$ symmetry. 
On the other hand, such a situation qualitatively changes once the ordering wave vectors lie on the low-symmetric points except for $\langle 100 \rangle$ and $\langle 110 \rangle$ lines as found in Figs.~\ref{fig: chi}(b)--\ref{fig: chi}(d); there is no constraint on the spiral plane owing to the arbitrariness of the DM vector direction. 
In other words, the proper-screw spiral state with the spiral plane perpendicular to $\bm{q}$ is possible, which is usually expected under the chiral point group like $O$ and $D_4$ rather than the polar one. 
For example, in the case of Fig.~\ref{fig: chi}(b), the spiral plane lies perpendicular to $\bm{D}_{\bm{q}}$ at $\bm{q}=(2\pi/5, 3\pi/20)$, where $\bm{D}_{\bm{q}}$ is given by $\bm{D}_{\bm{q}}=J^2_{\rm K}({\rm Im}[\chi^{yz}(\bm{q})], {\rm Im}[\chi^{zx}(\bm{q})])= J^2_{\rm K}(-0.00703, -0.05785)$; the spiral state is neither proper-screw nor cycloidal.
Such a situation also happens in EuNiGe$_3$, where the observed spiral state is characterized by a superposition of the proper-screw and cycloidal spiral waves~\cite{singh2023transition, matsumura2023distorted}. 

\section{Effective spin model and method}
\label{sec: Effective spin model and method}

The results in Sec.~\ref{sec: Effective spin interactions in itinerant electron systems} indicate that there is a possibility of realizing the hybrid SkX and anti-SkX when the ordering wave vectors lie on low-symmetric ones, which makes the direction of the DM vector arbitrary. 
We consider such a situation in order to investigate the stability of these unconventional SkXs in the ground state. 
For that purpose, we analyze an effective spin model of the Kondo lattice model in Eq.~(\ref{eq: Ham_krep})~\cite{hayami2021topological}, which is given by
\begin{align}
\label{eq: Ham}
\mathcal{H}^{\rm eff}=&-\sum_{\eta}  [J \bm{S}_{\bm{Q}_\eta} \cdot  \bm{S}_{-\bm{Q}_\eta} 
+ i \bm{D}_{\bm{Q}_\eta} \cdot (\bm{S}_{\bm{Q}_\eta} \times  \bm{S}_{-\bm{Q}_\eta}) \nonumber \\
&+ \sum_{\nu=x,y}\Gamma^\nu_{\bm{Q}_\eta} S^\nu_{\bm{Q}_\eta}S^\nu_{-\bm{Q}_\eta} ] 
-H\sum_i  S^z_i. 
\end{align}
This model is obtained by extracting the specific momentum-resolved interaction that gives the dominant contribution to the ground-state energy in Eq.~(\ref{eq:RKKYHam}). 
The first term represents the momentum-resolve interaction at wave vectors $\bm{Q}_\eta$, where $\eta$ is the index for the symmetry-related wave vectors. 
For the specific wave vectors, we choose $\pm \bm{Q}_1=\pm (Q_a, Q_b)$, $\pm \bm{Q}_2=\pm (-Q_b, Q_a)$, $\pm \bm{Q}_3=\pm (Q_a, -Q_b)$, and $\pm \bm{Q}_4=\pm (Q_b, Q_a)$ with $Q_a=13\pi/25$ and $Q_b=3\pi/25$ so that the ordering vectors are not on the high-symmetric $\langle 100 \rangle$ and $\langle 110 \rangle$ lines. 
It is noted that $\bm{Q}_1$--$\bm{Q}_4$ are connected by the fourfold rotational and mirror symmetries of the square lattice under the $C_{\rm 4v}$ point group.

At $\bm{Q}_1$--$\bm{Q}_4$, we consider the isotropic exchange interaction in the form of $J \bm{S}_{\bm{Q}_\eta} \cdot  \bm{S}_{-\bm{Q}_\eta} $, the DM interaction in the form of $i \bm{D}_{\bm{Q}_\eta} \cdot (\bm{S}_{\bm{Q}_\eta} \times  \bm{S}_{-\bm{Q}_\eta})$ with $\bm{D}_{\bm{Q}_\nu} = -\bm{D}_{-\bm{Q}_\nu}$, and the symmetric anisotropic exchange interaction in the form of $\Gamma^x_{\bm{Q}_\eta} S^x_{\bm{Q}_\eta}S^x_{-\bm{Q}_\eta}+\Gamma^y_{\bm{Q}_\eta} S^y_{\bm{Q}_\eta}S^y_{-\bm{Q}_\eta}$. 
The direction of the DM vector is $\bm{Q}_\eta$-dependent; we set $\bm{D}_{\bm{Q}_1}=D(-\cos \theta, \sin \theta)$ and other $\bm{D}_{\bm{Q}_\eta}$ in order to satisfy the polar symmetry~\cite{Yambe_PhysRevB.106.174437}. 
The symmetric anisotropic exchange interaction also has $\bm{Q}_\eta$ dependence; we set $(\Gamma^x_{\bm{Q}_1}, \Gamma^y_{\bm{Q}_1})=(\Gamma, 0)$ and other $(\Gamma^x_{\bm{Q}_\eta}, \Gamma^y_{\bm{Q}_\eta})$ to satisfy the polar symmetry, which also affects the stability of the SkX~\cite{amoroso2020spontaneous, yambe2021skyrmion, Hayami_PhysRevB.103.024439, Hayami_PhysRevB.103.054422, amoroso2021tuning, Hirschberger_10.1088/1367-2630/abdef9, Utesov_PhysRevB.103.064414, Wang_PhysRevB.103.104408, Kato_PhysRevB.104.224405,  Nickel_PhysRevB.108.L180411}. 
Although the magnitudes of the parameters $(J, D, \Gamma)$ in the model in Eq.~(\ref{eq: Ham}) are determined by the magnetic susceptibility in Eq.~(\ref{eq:RKKYHam}), we deal with them phenomenologically; we take $J=1$ as the energy unit of the model, and set $D=0.2$ and $\Gamma=0$ in Sec.~\ref{sec: Skyrmion crystal phases} or $\Gamma=0.1$ in Sec.~\ref{sec: Effect of symmetric anisotropic exchange interaction}. 
We ignore the effect of other symmetric anisotropic exchange interactions such as $S^x_{\bm{Q}_\eta}S^y_{-\bm{Q}_\eta}+S^y_{\bm{Q}_\eta}S^x_{-\bm{Q}_\eta}$ that arises from ${\rm Re}[\chi^{xy}(\bm{q})]$ in Eq.~(\ref{eq: suscep}) for simplicity. 
In addition, we introduce the second term in Eq.~(\ref{eq: Ham}), which represents the Zeeman term under an external magnetic field along the $z$ direction. 

The magnetic phase diagram at low temperatures is constructed by performing the simulated annealing from high temperatures $T_0=$1--5 for the spin model with the system size $N=50^2$ under the periodic boundary conditions, where $N$ represents the total number of sites.  
Starting from a random spin configuration, we gradually reduce the temperature as $T_{n+1}=0.999999 T_n$ to the final temperature $T=0.01$, where $T_n$ is the $n$th temperature. 
In each temperature, the spin is locally updated one by one following the standard Metropolis algorithm. 
When the temperature reaches the final temperature $T$, further Monte Carlo sweeps around $10^5$--$10^6$ are performed for measurements. 
The simulations independently run for different model parameters. 
In order to avoid the meta-stable solutions in the vicinity of the phase boundaries, the simulations from the spin configurations obtained at low temperatures are also performed. 

The spin and scalar spin chirality quantities are calculated to identify magnetic phases. 
The uniform magnetization along the field direction is given by 
\begin{align}
\label{eq: magnetization}
M^z =\frac{1}{N} \sum_i S_i^z.   
\end{align} 
The spin structure factor is given by 
\begin{align}
S_s(\bm{q})&=\sum_{\nu} S^{\nu\nu}_s(\bm{q}) \\
S^{\nu\nu}_s(\bm{q})&=\frac{1}{N}\sum_{ij}S_i^\nu S_j^\nu e^{i \bm{q} \cdot (\bm{r}_i -\bm{r}_j)}, 
\end{align}
for $\nu = x, y, z$. 
$\bm{r}_i$ represents the position vector at site $i$ and $\bm{q}$ represents the wave vector in the first Brillouin zone. 
The scalar spin chirality is given by 
\begin{align}
\chi^{\rm sc}&= \frac{1}{2 N} 
\sum_{i}
\sum_{\delta,\delta'= \pm1}
\delta \delta'
\bm{S}_{i} \cdot (\bm{S}_{i+\delta\hat{x}} \times \bm{S}_{i+\delta'\hat{y}}), 
\end{align}
where $\hat{x}$ ($\hat{y}$) represents a shift by lattice constant in the $x$ ($y$) direction. 
The scalar spin chirality is one of the signals to identify the SkX, i.e., $\chi^{\rm sc} \neq 0$ for the SkX.

\section{Skyrmion crystal phases}
\label{sec: Skyrmion crystal phases}

\begin{figure}[tb!]
\begin{center}
\includegraphics[width=1.0\hsize]{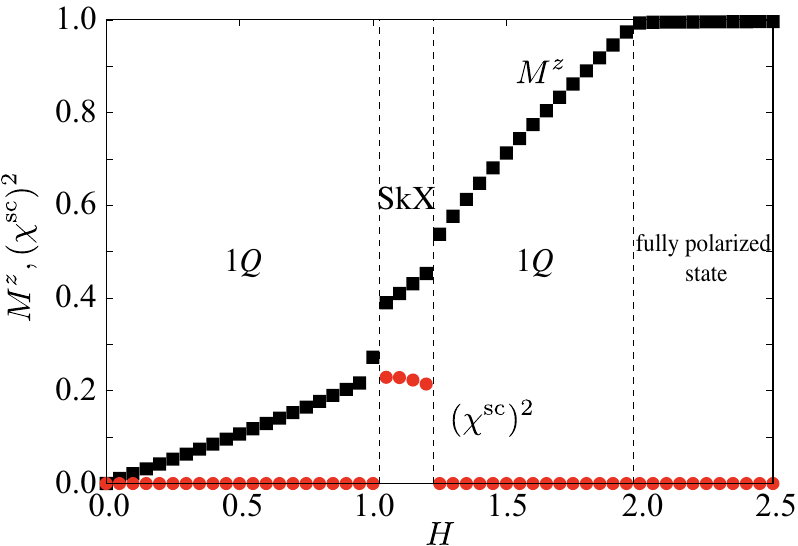} 
\caption{
\label{fig: mag_wo_ani} 
$H$ dependence of the $z$ component of the magnetization $M^z$ and the squared scalar chirality $(\chi^{\rm sc})^2$ for $\theta=0.4\pi$. 
The vertical dashed lines represent the phase boundaries between different magnetic phases. 
}
\end{center}
\end{figure}

We discuss the effect of the DM interaction at low-symmetric wave vectors on the stabilization of the SkX; we set $\Gamma=0$ and discuss its effect in Sec.~\ref{sec: Effect of symmetric anisotropic exchange interaction}. 
Figure~\ref{sec: Skyrmion crystal phases} shows the $H$ dependence of the $z$ component of the magnetization $M^z$ and the squared scalar spin chirality $(\chi^{\rm sc})^2$ at $\theta=0.4 \pi$. 
For $H=0$, the ground-state spin configuration corresponds to the single-$Q$ spiral (1$Q$) state, whose ordering wave vector is characterized by any of $\bm{Q}_1$--$\bm{Q}_4$. 
The spiral plane is determined so as to align perpendicular to $\bm{D}_{\bm{Q}_\eta}$, which results in the energy gain by the DM interaction. 
In the case of the $\bm{Q}_1$ ordering wave vector, the DM interaction with $\theta=0.4\pi$ is given by $\bm{D}_{\bm{Q}_1} \simeq D (-0.309, 0.951)$, which results in the spiral plane on (0.951,0.309). 
When the magnetic field is turned on, the spiral plane is gradually tilted to the plane perpendicular to the magnetic field to gain the energy by the Zeeman coupling. 
When $H$ reaches $1.05$, the 1$Q$ state is replaced by the SkX, which is characterized by the double-$Q$ spiral waves as detailed below; the scalar chirality becomes nonzero, as shown in Fig.~\ref{fig: mag_wo_ani}. 
Then, the SkX turns into the 1$Q$ state again by further increasing $H$, where the spiral plane is almost perpendicular to the field direction. 
Finally, the 1$Q$ state continuously changes into the fully polarized state.

\begin{figure}[tb!]
\begin{center}
\includegraphics[width=1.0\hsize]{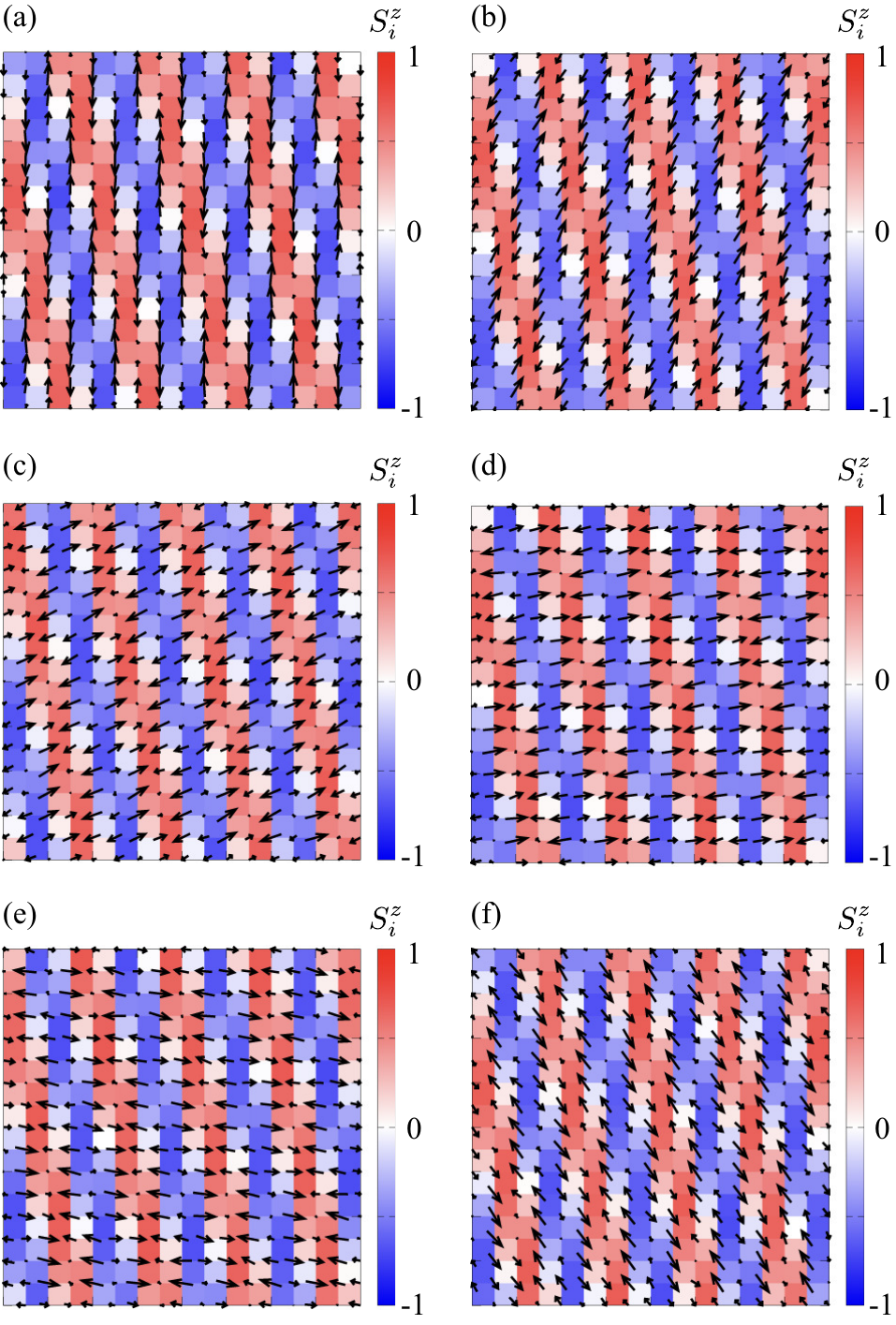} 
\caption{
\label{fig: Spin_1Q} 
Real-space spin configurations in the 1$Q$ state for $H=0$ at (a) $\theta=0$, (b) $\theta=0.16\pi$, (c) $\theta=0.36\pi$, (d) $\theta=0.44\pi$, (e) $\theta=0.56\pi$, and (f) $\theta=0.8\pi$. 
The arrows represent the direction of the in-plane spin moments and the color shows its $z$ component. 
}
\end{center}
\end{figure}

\begin{figure}[tb!]
\begin{center}
\includegraphics[width=1.0\hsize]{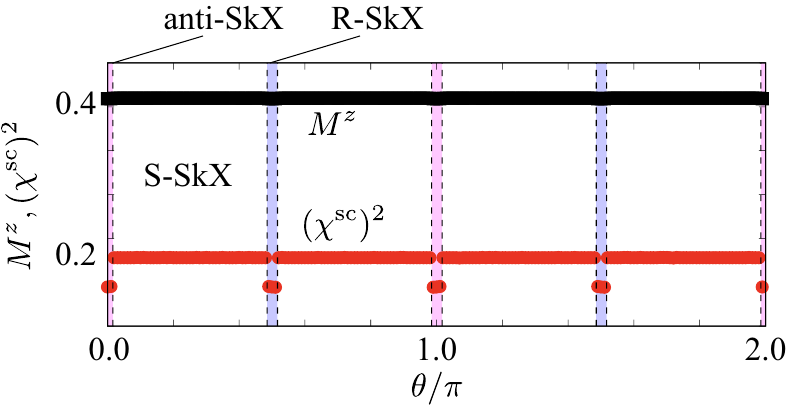} 
\caption{
\label{fig: mag_Dthetadep} 
$\theta$ dependence of $M^z$ and $(\chi^{\rm sc})^2$ at $H=1.1$. 
The vertical dashed lines represent the phase boundaries between different magnetic phases. 
The regions in blue (pink) represent the R-SkX (anti-SkX) phase, while those in white represent the S-SkX. 
}
\end{center}
\end{figure}

\begin{figure}[tb!]
\begin{center}
\includegraphics[width=1.0\hsize]{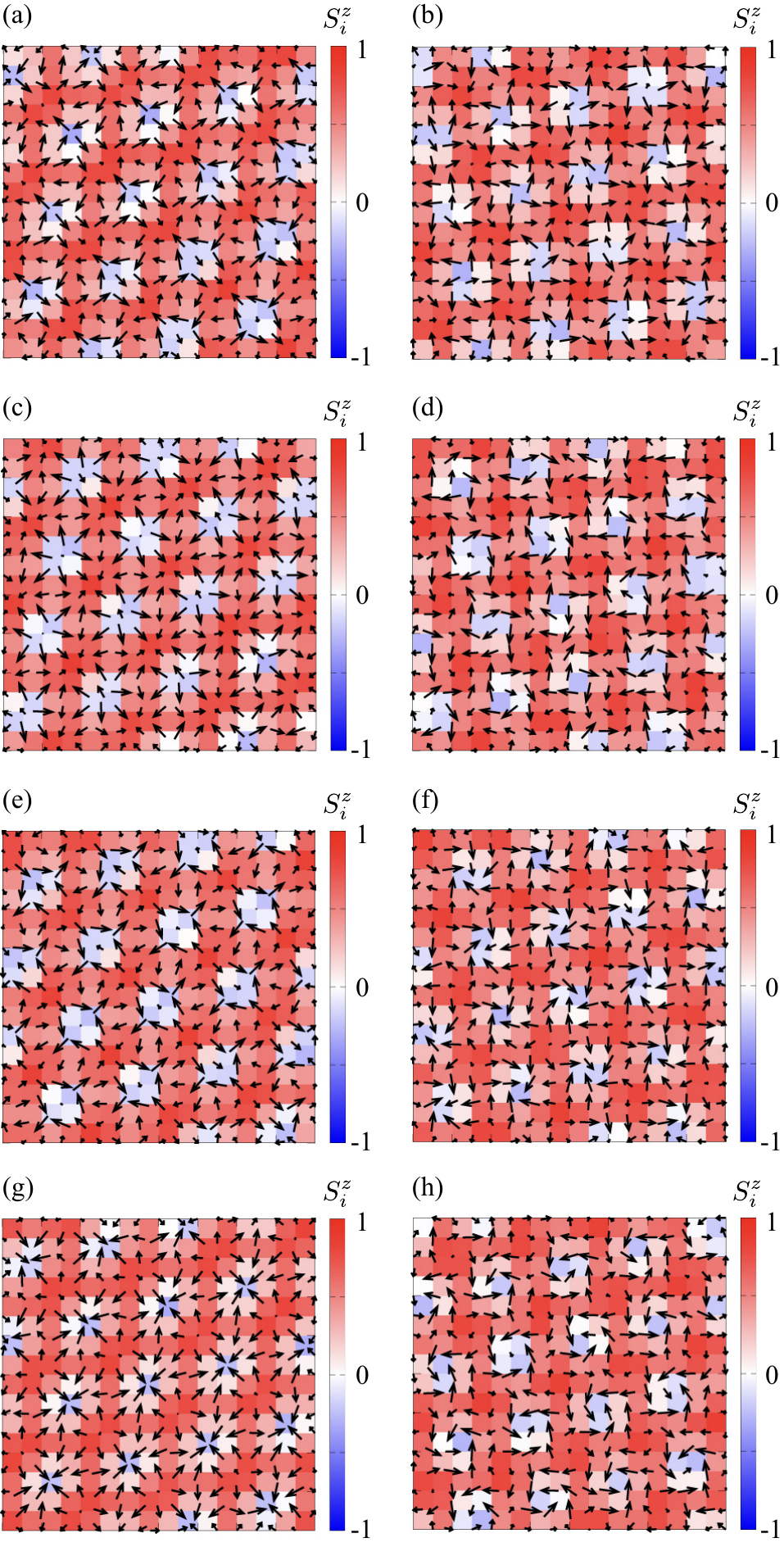} 
\caption{
\label{fig: Spin_SkX} 
Real-space spin configurations in the SkX for $H=1.1$ at (a) $\theta=0$, (b) $\theta=0.25\pi$, (c) $\theta=0.5\pi$, (d) $\theta=0.75\pi$, (e) $\theta=\pi$, (f) $\theta=1.25\pi$, (g) $\theta=1.5\pi$, and (h) $\theta=1.75\pi$.
The spin configurations in (a) and (e) correspond to the anti-SkX, those in (b), (d), (f), and (h) correspond to the S-SkX and those in (c) and (g) correspond to the R-SkX,  
The arrows represent the direction of the in-plane spin moments and the color shows its $z$ component. 
}
\end{center}
\end{figure}

\begin{figure}[tb!]
\begin{center}
\includegraphics[width=1.0\hsize]{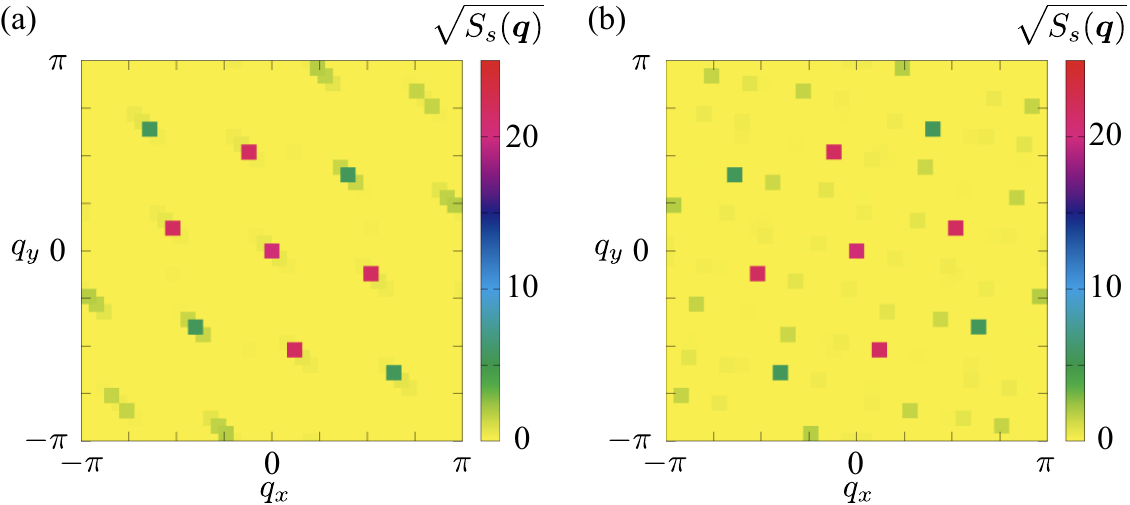} 
\caption{
\label{fig: Sq} 
The square root of the spin structure factor in the anti-SkX at (a) $\theta=0$ and (b) S-SkX at $\theta=0.25\pi$. 
The spin structure factor of the R-SkX is the same as that in the anti-SkX. 
}
\end{center}
\end{figure}

The above phase sequence against $H$ is independent of $\theta$ within the discretized data in Fig.~\ref{fig: mag_wo_ani}. 
Meanwhile, we find two characteristic features in terms of the $\theta$ dependence. 
One is the $\theta$-dependent spiral plane in the 1$Q$ state. 
We show the real-space spin configurations at $H=0$ for different $\theta$ in Figs.~\ref{fig: Spin_1Q}(a)--\ref{fig: Spin_1Q}(f), where the ordering wave vector is chosen as $\bm{Q}_1$. 
One finds that the spiral plane of the 1$Q$ state changes according to $\theta$, which is understood from the fact that the spiral plane is determined by the direction of the DM vector, as discussed above. 
Thus, the cycloidal spiral state with the spiral plane parallel to the wave vector, which usually appears under the polar symmetry, is not necessarily realized once the ordering wave vectors lie at the low-symmetric ones. 
In other words, the proper-screw spiral state with the spiral plane perpendicular to the wave vector can be also induced at the low-symmetric wave vectors. 
Indeed, such a tendency has been found in the tetragonal polar magnet EuNiGe$_3$, where the ordering wave vector lies at the low-symmetric position~\cite{singh2023transition, matsumura2023distorted}. 
For almost all of $\theta$, the spiral plane is neither parallel nor perpendicular to the ordering wave vector. 

The other characteristic feature of the $\theta$ dependence appears in nonzero $H$. 
We show the $\theta$ dependence of $M^z$ and $(\chi^{\rm sc})^2$ at $H=1.1$ in Fig.~\ref{fig: mag_Dthetadep}, where the SkX with nonzero $(\chi^{\rm sc})^2$ appears irrespective of $\theta$. 
Intriguingly, different three types of SkXs are realized depending on $\theta$: the anti-SkX for $\theta \simeq 0$ and $\pi$, R-SkX for $\theta \simeq \pi/2, 3\pi/2$, and S-SkX for other $\theta$. 
We show the real-space spin configurations for several different $\theta$ in Fig.~\ref{fig: Spin_SkX}; the spin configurations in Figs.~\ref{fig: Spin_SkX}(a) and \ref{fig: Spin_SkX}(e) correspond to that in the anti-SkX, the spin configurations in Figs.~\ref{fig: Spin_SkX}(c) and \ref{fig: Spin_SkX}(g) correspond to that in the R-SkX, and the spin configurations in Figs.~\ref{fig: Spin_SkX}(b), \ref{fig: Spin_SkX}(d), \ref{fig: Spin_SkX}(f), and \ref{fig: Spin_SkX}(h) correspond to the S-SkX. 
The anti-SkX and the R-SkX are characterized by a superposition of double-$Q$ spiral waves at $\bm{Q}_2$ and $\bm{Q}_3$ [Fig.~\ref{fig: Sq}(a)], while the S-SkX is characterized by that at $\bm{Q}_1$ and $\bm{Q}_2$ [Fig.~\ref{fig: Sq}(b)]. 
Reflecting the direction of the DM vector, the SkXs with various values of helicity, i.e., the hybrid SkXs, are realized for almost all of $\theta$.

The above results indicate that there are two possibilities for constructing the double-$Q$ SkX in the tetragonal system. 
One is the case where the constituent double-$Q$ ordering wave vectors are connected by the mirror symmetry and the other is the case where the constituent double-$Q$ ordering wave vectors are connected by the fourfold rotational symmetry; the former leads to the R-SkX (or anti-SkX) and the latter leads to the S-SkX, where the alignment of the skyrmion core is different from each other, as shown by the real-space spin configurations in Fig.~\ref{fig: Spin_SkX}. 
The choice of two alignments is determined by the direction of the DM vector; from the simulation results, the R-SkX (or anti-SkX) tends to be stabilized when $\bm{D}_{\bm{Q}_\eta}$ is almost characterized by only one component, such as $\bm{D}_{\bm{Q}_\eta} \simeq (D, 0, 0)$ and $\bm{D}_{\bm{Q}_\eta} \simeq (0, D, 0)$.
From an energetic viewpoint, the R-SkX (or anti-SkX) is almost degenerate to the S-SkX, which implies that the lower-energy state might be accidentally determined. 
Thus, the contribution from high-harmonic wave vectors that are not taken into account in the present model plays an important role in enhancing either of the SkXs~\cite{Hayami_doi:10.7566/JPSJ.89.103702, Hayami_PhysRevB.105.174437, hayami2022multiple}.  
For example, the contributions from the wave vectors $\bm{Q}_2+ \bm{Q}_3= (Q_a-Q_b, Q_a-Q_b)$ and $\bm{Q}_2- \bm{Q}_3=(-Q_a-Q_b, Q_a+Q_b)$ tend to stabilize the R-SkX (or anti-SkX), while those from $\bm{Q}_1+\bm{Q}_2=(Q_a-Q_b, Q_a+Q_b)$ and $\bm{Q}_1-\bm{Q}_2=(Q_a+Q_b, -Q_a+Q_b)$ tend to stabilize the S-SkX. 

The different choices of the constituent ordering wave vectors in the SkXs result in the sign change of the scalar spin chirality. 
In the S-SkX, the scalar spin chirality always takes negative values, since the constituent spiral waves at $\bm{Q}_1$ and $\bm{Q}_2$ are related by the fourfold rotational symmetry; the spin texture around each skyrmion has the skyrmion number of $-1$. 
Meanwhile, the situation changes in the R-SkX and anti-SkX, whose ordering wave vectors are related by the vertical mirror symmetry. 
For $\theta =0$ ($\theta=\pi$), $\bm{D}_{\bm{Q}_2}$ are related to $\bm{D}_{\bm{Q}_3}$ by the rotation $\pi/2$ ($-\pi/2$), which indicates that the superposition of the spiral waves at $\bm{Q}_2$ and $\bm{Q}_3$ leads to the SkX with the skyrmion number of $-1$ ($+1$). 
Indeed, the in-plane component of the spins surrounding the skyrmion core form the vortex (antivortex) winding for the R-SkX (anti-SkX), as shown by the real-space spin configuration in Figs.~\ref{fig: Spin_SkX}(c) and \ref{fig: Spin_SkX}(g) [Figs.~\ref{fig: Spin_SkX}(a) and \ref{fig: Spin_SkX}(e)]. 
Thus, the anti-SkX with the positive skyrmion number is possible even in polar magnets when the low-symmetric wave vectors become the ordering wave vectors.

\section{Effect of symmetric anisotropic exchange interaction}
\label{sec: Effect of symmetric anisotropic exchange interaction}

\begin{figure}[tb!]
\begin{center}
\includegraphics[width=1.0\hsize]{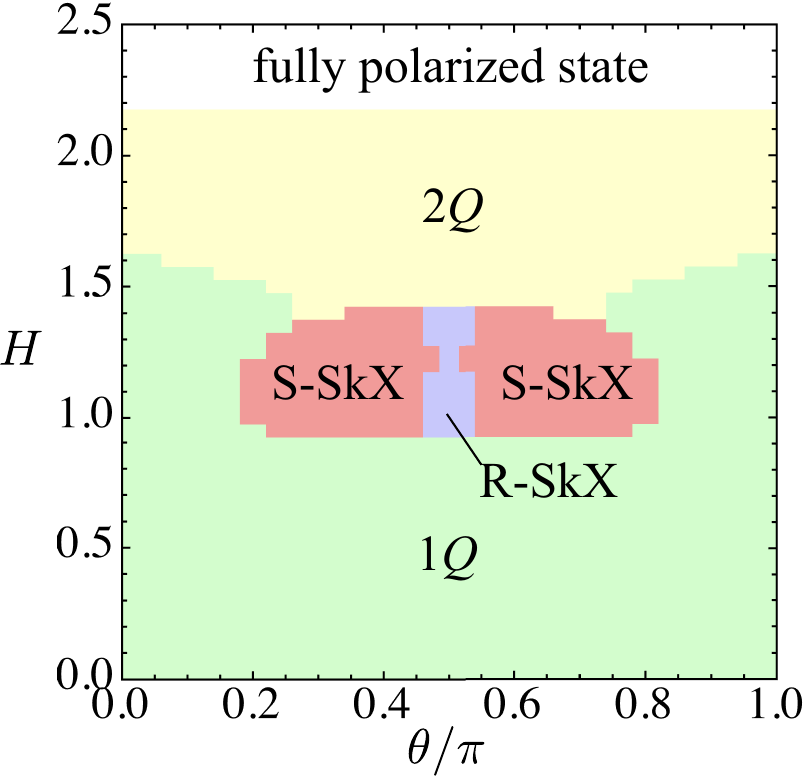} 
\caption{
\label{fig: PD} 
Magnetic phase diagram of the model in Eq.~(\ref{eq: Ham}) with changing $\theta$ and $H$ at $\Gamma=0.1$. 
}
\end{center}
\end{figure}

\begin{figure}[tb!]
\begin{center}
\includegraphics[width=1.0\hsize]{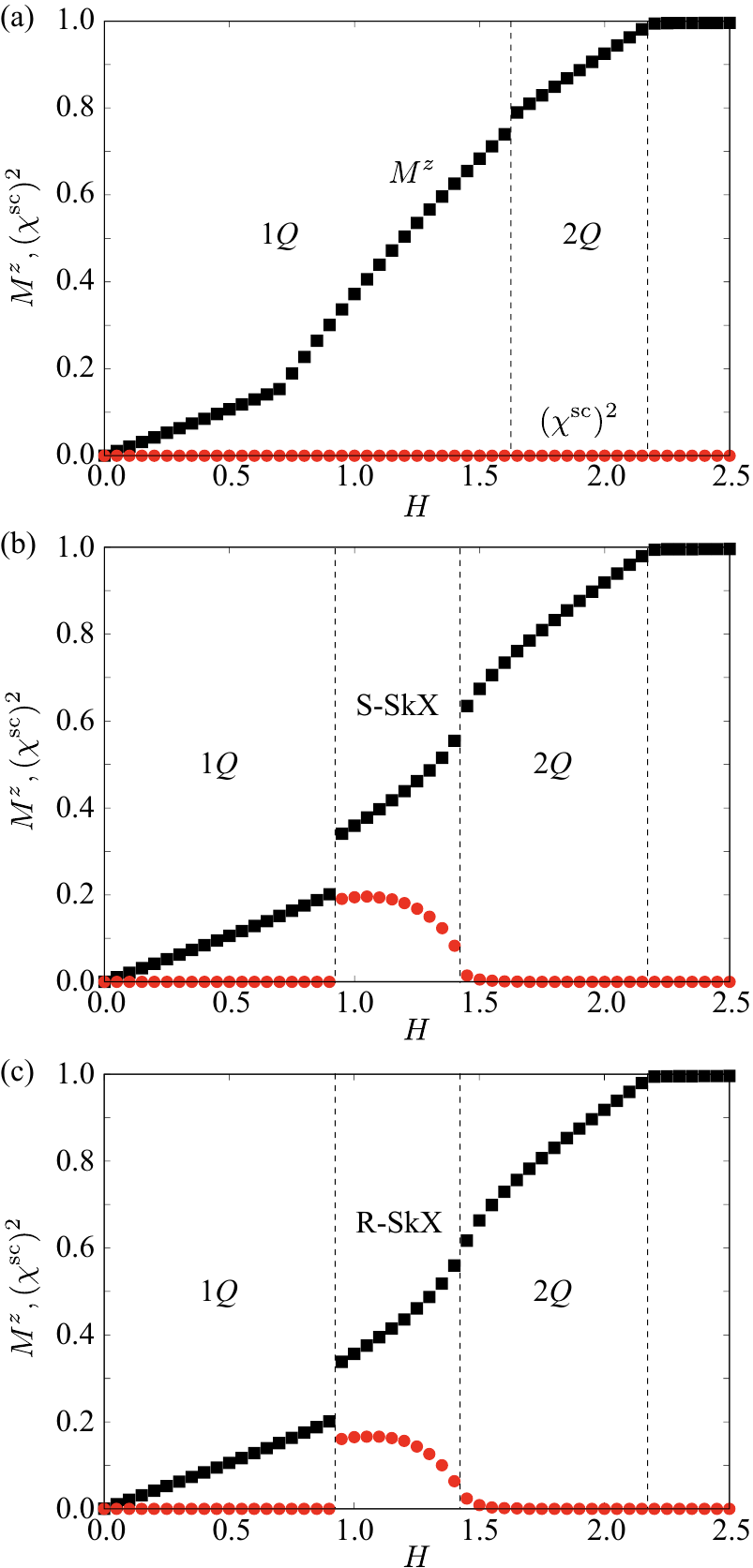} 
\caption{
\label{fig: mag} 
$H$ dependence of $M^z$ and $(\chi^{\rm sc})^2$ for (a) $\theta=0$, (b) $\theta=0.4\pi$, and (c) $\theta=0.5\pi$. 
The vertical dashed lines represent the phase boundaries between different magnetic phases. 
}
\end{center}
\end{figure}

\begin{figure}[tb!]
\begin{center}
\includegraphics[width=0.9\hsize]{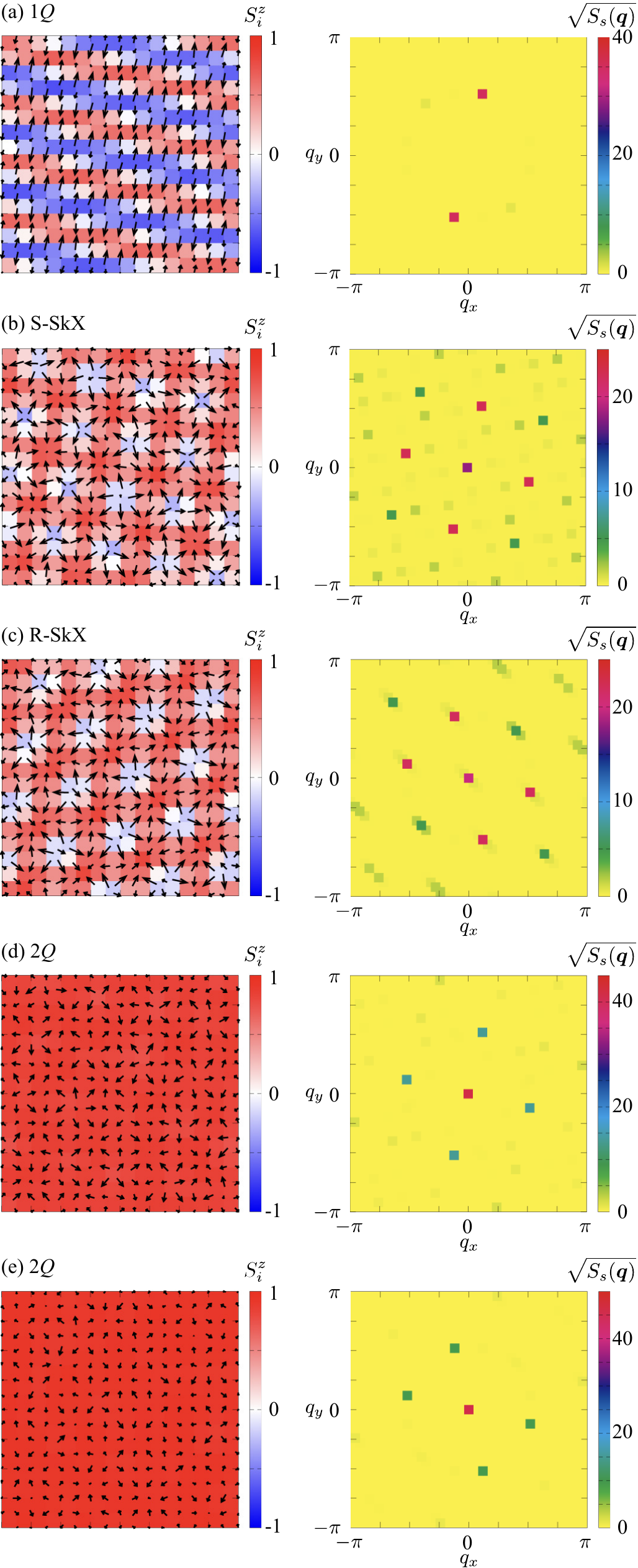} 
\caption{
\label{fig: Spin} 
(Left panel) Real-space spin configurations in 
(a) the 1$Q$ state at $\theta=0.4\pi$ and $H=0$, 
(b) the S-SkX at $\theta=0.4\pi$ and $H=1$, 
(c) the R-SkX at $\theta=0.5\pi$ and $H=1.1$, 
(d) the 2$Q$ state at $\theta=0.4\pi$ and $H=1.8$, and 
(e) another 2$Q$ state at $\theta=0.4\pi$ and $H=2$. 
The arrows represent the direction of the in-plane spin moments and the color shows its $z$ component. 
(Right panel) The square root of the spin structure factor corresponding to the left panel.
}
\end{center}
\end{figure}

In this section, we consider the effect of the symmetric anisotropic exchange interaction $\Gamma$, which can be the origin of the hybrid SkX and anti-SkX even without the DM interaction~\cite{Hayami_doi:10.7566/JPSJ.89.103702, yao2020controlling, Hayami_PhysRevB.105.104428, Kong_PhysRevB.109.014401}, on the stability of the SkX in Sec.~\ref{sec: Skyrmion crystal phases}. 
We set $\Gamma=0.1$. 

Figure~\ref{fig: PD} shows the magnetic phase diagram in the plane of $\theta$ and $H$. 
Compared to the result in Fig.~\ref{fig: mag_wo_ani}, where the SkX is stabilized for $1.03 \lesssim H \lesssim 1.23$, the region of the SkX becomes large for $0.2\pi \lesssim \theta \lesssim 0.8 \pi$, while that vanishes for $0  \lesssim \theta \lesssim 0.2 \pi$ and $0.8\pi \lesssim \theta \lesssim  \pi$. 
Thus, the anti-SkX no longer appears in the phase diagram for $\Gamma=0.1$. 
We show the $H$ dependence of $M^z$ and $(\chi^{\rm sc})^2$ for $\theta=0$ in Fig.~\ref{fig: mag}(a), $\theta=0.4\pi$ in Fig.~\ref{fig: mag}(b), and $\theta=0.5 \pi$ in Fig.~\ref{fig: mag}(c). 

In the low-field region, the 1$Q$ state is stabilized irrespective of $\theta$, although its direction of the spiral plane depends on $\theta$.  
The real-space spin configuration and the spin structure factor of the 1$Q$ state are shown in the left and right panels of Fig.~\ref{fig: Spin}(a), respectively. 
Since $\Gamma=0.1$ tends to favor the oscillation in terms of the $x$ ($y$) spin component for $\bm{Q}_1$ and $\bm{Q}_3$ ($\bm{Q}_2$ and $\bm{Q}_4$), the spiral plane is tilted from the plane perpendicular to $\bm{D}_{\bm{Q}_\eta}$ to gain the energy by $\Gamma$. 

In the intermediate-field region, the stability region of the SkX is enhanced for $0.2\pi \lesssim \theta \lesssim 0.8 \pi$, while it is suppressed for $0  \lesssim \theta \lesssim 0.2 \pi$ and $0.8\pi \lesssim \theta \lesssim  \pi$; the real-space spin configurations and the spin structure factors of the S-SkX and R-SkX are shown in the left and right panels of Figs.~\ref{fig: Spin}(b) and \ref{fig: Spin}(c), respectively. 
This stability tendency is understood from the effect of $\Gamma$. 
For example, for $\theta=0.5\pi$, the DM interaction at $\bm{Q}_1$ tends to favor the spiral wave in the $xz$ plane, while the symmetric anisotropic exchange interaction at $\bm{Q}_1$ tends to favor the $x$ spin oscillation. 
Thus, the effects of $D$ and $\Gamma$ are cooperative in enhancing the stability of the spiral wave in the $xz$ plane. 
A similar tendency also holds for other $\bm{Q}_\eta$; for example in the case of $\bm{Q}_4$, both $D$ and $\Gamma$ tend to favor the spiral wave in the $yz$ plane. 
On the other hand, for $\theta=0$, $D$ and $\Gamma$ lead to different spiral states; $D$ at $\bm{Q}_1$ tends to favor the spiral state in the $yz$ plane and $\Gamma$ at $\bm{Q}_1$ tends to favor the spiral state in the $xz$ plane. 
This indicates frustration between $D$ and $\Gamma$, which avoids the stabilization of the SkX in the region near $\theta =0 $ and $\theta=\pi$. 
It is noted that the opposite tendency can happen when we consider $\Gamma=-0.1$ so that the spiral state in the $yz$ ($xz$) plane is favored at $\bm{Q}_1$ ($\bm{Q}_4$); the anti-SkX remains stable, whereas the R-SkX vanishes. 
In the end, the relative relationship between $D$ and $\Gamma$ is important whether the SkX appears or not. 

In the high-field region, the double-$Q$ (2$Q$) state appears instead of the 1$Q$ state irrespective of $\theta$.  
The spin configuration of the 2$Q$ state is almost characterized by the in-plane spin modulations at $\bm{Q}_3$ and $\bm{Q}_4$ or $\bm{Q}_2$ and $\bm{Q}_3$, as shown by the real-space spin configurations and spin structure factors in Figs.~\ref{fig: Spin}(d) and \ref{fig: Spin}(e). 
Since the energy in the 2$Q$ state with $\bm{Q}_3$ and $\bm{Q}_4$ is almost the same as that in the 2$Q$ state with $\bm{Q}_2$ and $\bm{Q}_3$, it is difficult to distinguish them in the present phase diagram; the additional effect such as $S^x_{\bm{Q}_\eta}S^y_{-\bm{Q}_\eta}+S^y_{\bm{Q}_\eta}S^x_{-\bm{Q}_\eta}$ will lift such a degeneracy. 
The 2$Q$ state continuously turns into the fully polarized state when the magnetic field increases, as shown in Figs.~\ref{fig: mag}(a)--\ref{fig: mag}(c).

\section{Summary}
\label{sec: Summary}

To summarize, we have investigated the role of the DM interaction at low-symmetric wave vectors. 
We have analyzed the effective spin model on the polar square lattice, which is derived from the Kondo lattice model in the weak-coupling regime, by performing the simulated annealing.  
We have found that the direction of the DM vector affects the formation of the SkXs as well as the helicity of the spiral wave. 
We have shown that the R-SkX is realized when the DM vector lies in the $\langle 100 \rangle$ direction, while the S-SkX is realized for other cases. 
Furthermore, we have shown that the anti-SkX is also realized depending on the direction of the DM vector, which provides another root to realize the anti-SkX even under polar symmetry. 
The present results indicate that the low-symmetric ordering wave vectors become a source of inducing further intriguing SkXs. 

The present situation also holds for other noncentrosymmetric systems. 
For example, the system with the chiral-type DM interaction under the $D_4$ (422) point group, which usually favors the Bloch SkX, can also lead to the hybrid SkX and anti-SkX once the ordering wave vectors lie at the low-symmetric ones. 
In addition, one can expect such generations of the hybrid SkX and anti-SkX induced by the DM interaction at low-symmetric wave vectors in centrosymmetric systems with the lack of local inversion symmetry~\cite{Hayami_PhysRevB.105.014408, lin2021skyrmion, Hayami_PhysRevB.105.184426, hayami2022square}. 
In this case, the sublattice-dependent DM interaction becomes the origin of the above unconventional SkXs.

\begin{acknowledgments}
The author thanks J. S. White and D. Singh for fruitful discussions. 
This research was supported by JSPS KAKENHI Grants Numbers JP21H01037, JP22H04468, JP22H00101, JP22H01183, JP23H04869, JP23K03288, and by JST PRESTO (JPMJPR20L8) and JST CREST (JPMJCR23O4).  
Parts of the numerical calculations were performed in the supercomputing systems in ISSP, the University of Tokyo.
\end{acknowledgments}

\appendix

\bibliographystyle{apsrev}
\bibliography{../ref.bib}
\end{document}